\documentclass[journal]{IEEEtran}
\usepackage{graphicx}
\usepackage[cmex10]{amsmath}
\usepackage{float}
\usepackage{booktabs,multirow}

\usepackage{color}

\begin{document}

\title{Numerical Fokker-Planck study of stochastic write error slope in spin torque switching}

\author{Yunkun Xie, Behtash Behin-Aein, Avik W. Ghosh
\thanks{Yunkun Xie is with the Charles L. Brown Department of Electrical and Computer Engineering, University of Virginia, Charlottesville,
VA, 22904 USA (e-mail: yx3ga@virginia.edu).}
\thanks{Behtash Behin-Aein is with GLOBALFOUNDRIES, Sunnyvale, CA, 94085, USA (email: behtash.behin-aein@globalfoundries.com).}
\thanks{Avik W. Ghosh is with the Charles L. Brown Department of Electrical and Computer Engineering and Department of Physics, University of Virginia, Charlottesville,
VA, 22904 USA (e-mail: ag7rq@virginia.edu).}}

\markboth{Journal of \LaTeX\ Class Files,~Vol.~13, No.~9, September~2014}%
{Shell \MakeLowercase{\textit{et al.}}: Bare Demo of IEEEtran.cls for Journals}

\maketitle

\begin{abstract}
This paper analyzes write errors in spin torque switching due to thermal fluctuations in a system with Perpendicular Magnetic Anisotropy (PMA). Prior analytical and numerical methods are summarized, a physics based Fokker-Planck Equation (FPE) chosen for its computational efficiency and broad applicability to all switching regimes. The relation between write error slope and material parameters is discussed in detail to enable better device engineering and optimization. Finally a 2D FPE tool is demonstrated that extends the applicability of FPE to write error in non PMA systems with built-in asymmetry.
\end{abstract}

\begin{IEEEkeywords}
PMA, spin transfer torque, 2D Fokker-Planck, write error rate.
\end{IEEEkeywords}

\IEEEpeerreviewmaketitle

\section{Introduction}

\IEEEPARstart{O}{ver} the past decade, emerging spintronics and nanomagnetic devices have attracted a lot of attention due to their versatility, scalability and energy efficiency. Part of the excitement stems from the discovery and experimental demonstration of spin transfer torque (STT) effect\cite{slonczewski1996,berger1996}. Compared to traditional switching schemes using external magnetic fields, STT provides a scalable solution to manipulate the magnetization of a nano-sized magnet.  Several direct applications of STT such as spin transfer torque based magnetic random access memory (STT-MRAM) and spin torque oscillator (STO) for microwave sources and pattern recognition have been proposed and experimentally demonstrated\cite{hosomi2005,kiselev2003}. A common issue accompanying magnetic switching is its susceptibility to thermal noise. At room temperature the magnetic switching under STT reacts to thermal fluctuations and often results in a distribution of switching currents and delays. Some applications can explicitly utilize these thermal fluctuations, such as random number generators in spin dice\cite{Fukushima2014} or stochastic simulation of neuromorphic behavior\cite{Vincent2014}. In other applications such as STT based memory, the stochastic nature can cause read/write errors and need to be accounted for. Particularly in the case of write operation in STTRAMs, increasing the applied current or switching time can effectively reduce the write error rate but both quantities are limited by reliability, endurance and overall performance metrics. It is therefore important to develop a holistic understanding of the overall energy-delay-reliability tradeoffs in STT based memory\cite{munira2012}.

\section{Switching Regions and Analytical Models}
Discussions on thermal effects in spin torque switching usually fall into two switching regions set by the ratio between injected current $I$ and critical current $I_c$. The supercritical $I\gg I_c$ regime is called current dominated region while the subcritical regime $I\ll I_c$ is referred to as thermal region. This division not only differentiates between switching behavior in different regimes but also allows separate approximations that allow simple physics-based analytical models in the two regimes. Although these analytical equations capture switching features in their individual regions, they typically do not allow a `smooth' transition between them and often encounter mathematical singularities at the transition. Recently there have been efforts at formulating a brute force mathematical transition between the analytical equations\cite{Vincent2015}. Such a scheme offers a simple fix to the discontinuity of the analytical equations but lacks physical insights into the switching behavior at transition. The alternative is to use numerical solution to avoid mathematical approximations and take into account physical parameters at the same time. Numerical methods are quite universal and not limited to a specific region but are in general much less computationally efficient than analytical approaches. Two physics based numerical approaches are discussed in the next section, out of which we argue that the Fokker-Planck approach is more practical. Note that all equations to be discussed are based on macrospin approximation. In reality STT switching can involve complications like sub-volume effects \cite{Sun2011} or edge effects\cite{Song2015}. While it is crucial to understand those effects, accounting for both non-macrospin effects and thermal effects can be computationally challenging in simulations. Thus macrospin model is still valuable because one can typically approximate those complicated effects with effective parameters that are good enough to draw physical insights and at the same time interface with practical device or circuit simulations.

\subsection{Current dominated regime}
In the current dominated supercritical current regime, Sun's approach \cite{Sun2004} is very popular due to its simplicity. In Sun's equation, the probability of switching is a double exponential function of current and pulse width:

\begin{equation}
P_{sw}=\exp\left\{-4\Delta\exp\left[-2\tau(i-1)\right]\right\}
\label{eq:Sun}
\end{equation}

where $i,\tau,\Delta$ are scaled quantities defined as:

\begin{equation}
\begin{split}
i&=\frac{I}{I_c},\ \ \ \tau=\frac{t}{\tau_D},\ \ \ \Delta=\frac{\mu_0 H_kM_s\Omega}{2k_BT}\\
I_c&=\frac{2\alpha q}{\eta\hbar}\mu_0H_kM_s\Omega,\ \ \tau_D=\frac{1+\alpha^2}{\alpha\gamma\mu_0 H_k},\ \ H_k=\frac{2K_u}{\mu_0M_s}
\label{eq:parameters}
\end{split}
\end{equation}

with $I_c$ the critical current, $\Delta$ the thermal stability factor. $M_s$ is the saturation magnetization. $\eta$ is the spin polarization of the injected current. $\Omega$ is the volume of the magnet. $K_u$ and $H_k$ are anisotropy constant and anisotropy field respectively. $\alpha$ is the magnetic damping coefficient and $T$ is the temperature. $\mu_0$ is the permeability constant. $k_B$ is the Boltzmann constant. $\gamma$ is the gyromagnetic ratio.

The physical picture described by Sun's equation is a thermally disturbed magnetization that is switched by an overdrive current in a {\it deterministic} way. In other words, Sun's equation considers the equilibrium thermal distribution of initial magnetization before switching, but once the current is applied thermal noise is turned off. An overdrive current $I \gg I_c$ is also a necessary condition for the switching time approximation implicit in Sun's equation\cite{Sun2004}. A more accurate analytical description is obtained by solving Eq. \ref{eq:1D_FP} approximately in the overdrive regime (analytical 1D Fokker-Planck solution for PMA system)\cite{Butler2012}: 

\begin{equation}
 P_{sw}=\exp\left\{-\frac{\pi^2\Delta}{4}\frac{i-1}{i\exp\left[2\tau(i-1)\right]-1}\right\}
\label{eq:AFP}
\end{equation}

Eq. \ref{eq:AFP} also requires the overdrive condition $I\gg I_c$. Compared to Sun's equation the analytical FPE includes thermal noise during the switching process to some extent.

Despite the different conditions during switching underlying Eq. \ref{eq:Sun} and Eq. \ref{eq:AFP}, their difference is expected to be small in the operative supercritical regime. This is because a large driven spin torque will diminish any thermal effect during the switching process and makes the outcomes of the equations quite similar.  On the other hand, neither Sun's equation nor the analytical FPE correctly describes the intermediate or subcritical regime where the applied current is close to or below the critical current.

\subsection{Thermal regime}
In the thermal region $I\ll I_c$, switching happens due to thermal agitation. The switching rate can be described by an empirical thermal transition model:

\begin{equation}
 P_{sw}=1-\exp\left\{-tf_0\exp\left[-\Delta(1-i)^\beta\right]\right\}
 \label{eq:thermal}
\end{equation}

where $f_0$ is the empirical attempt frequency set between $10^{9}\sim10^{10}\,\mathrm{Hz}$ to describe experimental data in most magnets. The term $\Delta(1-i)^\beta$ represents an effective scaled energy barrier between the two stable states, in this case the two orientations along the easy axis. Parameter $\beta=1$ is for in-plane systems \cite{Li2004} and $\beta=2$ is for PMAs. In the thermal regime, the current contributes to switching by reducing the effective barrier. In \cite{Butler2012} Butler {\it et al} also obtained an analytical FPE solution in the thermal region when $I/I_c\ll1$:

\begin{equation}
 \begin{split}
  P_{sw}=&1-\exp\left\{-tf_0\exp\left[-\Delta(1-i)^2\right]\right\}\\
  f_0&=\frac{1}{\tau_D}\sqrt{\frac{\Delta}{\pi}\left(1-i\right)^2\left(1+i\right)}
 \end{split}
\label{eq:AFP_thermal}
\end{equation}

Notice that Eq. \ref{eq:AFP_thermal} is consistent with the empirical model except for the attempt frequency $f_0$ which is explicitly expressed in the analytical FPE solution.

\section{Numerical methods}

\subsection{Stochastic Landau-Lifshitz-Gilbert Equation}
The phenomenological Landau-Lifshitz-Gilbert (LLG) equation describes dynamics of the normalized magnetization $\mathbf{m}=\mathbf{M}/M_s$ determined by the torque from effective magnetic field $\mathbf{H}_\mathrm{eff}$, magnetic damping  and Slonczewski spin torque $\mathbf{L}_{\mathrm{STT}}$ in the case of STT switching. Thermal noise is modeled as a gaussian noisy field $\mathbf{h}_\mathrm{th}$.

\begin{equation}
\begin{split}
 \left(1+\alpha^2\right)\frac{d\mathbf{m}}{dt}&=\mathbf{L}_\mathrm{prec}+\mathbf{L}_\mathrm{damp}+\mathbf{L}_\mathrm{STT}\IEEEyesnumber\\
 \mathbf{L}_\mathrm{prec} &= -\mu_0\gamma\mathbf{m}\times\mathbf{H}_\mathrm{eff}\\
 \mathbf{L}_\mathrm{damp} &=-\alpha\mu_0\gamma\mathbf{m}\times\left(\mathbf{m}\times\mathbf{H}_\mathrm{eff}\right) \\ 
  \mathbf{L}_\mathrm{STT}&=-\frac{\mu_BI\eta}{q\Omega M_s}\mathbf{m}\times\left(\mathbf{m}\times\mathbf{I}_s\right)\\
  \mathbf{H}_\mathrm{eff}&=\mathbf{H}_\mathrm{anis}+\mathbf{H}_\mathrm{ext}+\mathbf{h}_\mathrm{th} \\
  &\mathbf{h}_\mathrm{th} = \sqrt{\frac{2\alpha k_BT}{\mu_0\gamma\Omega M_s}}\mathbf{G}
  \end{split}
    \label{eq:LLGS}
\end{equation}

\begin{figure}[H]
\centering
\includegraphics[width=7cm]{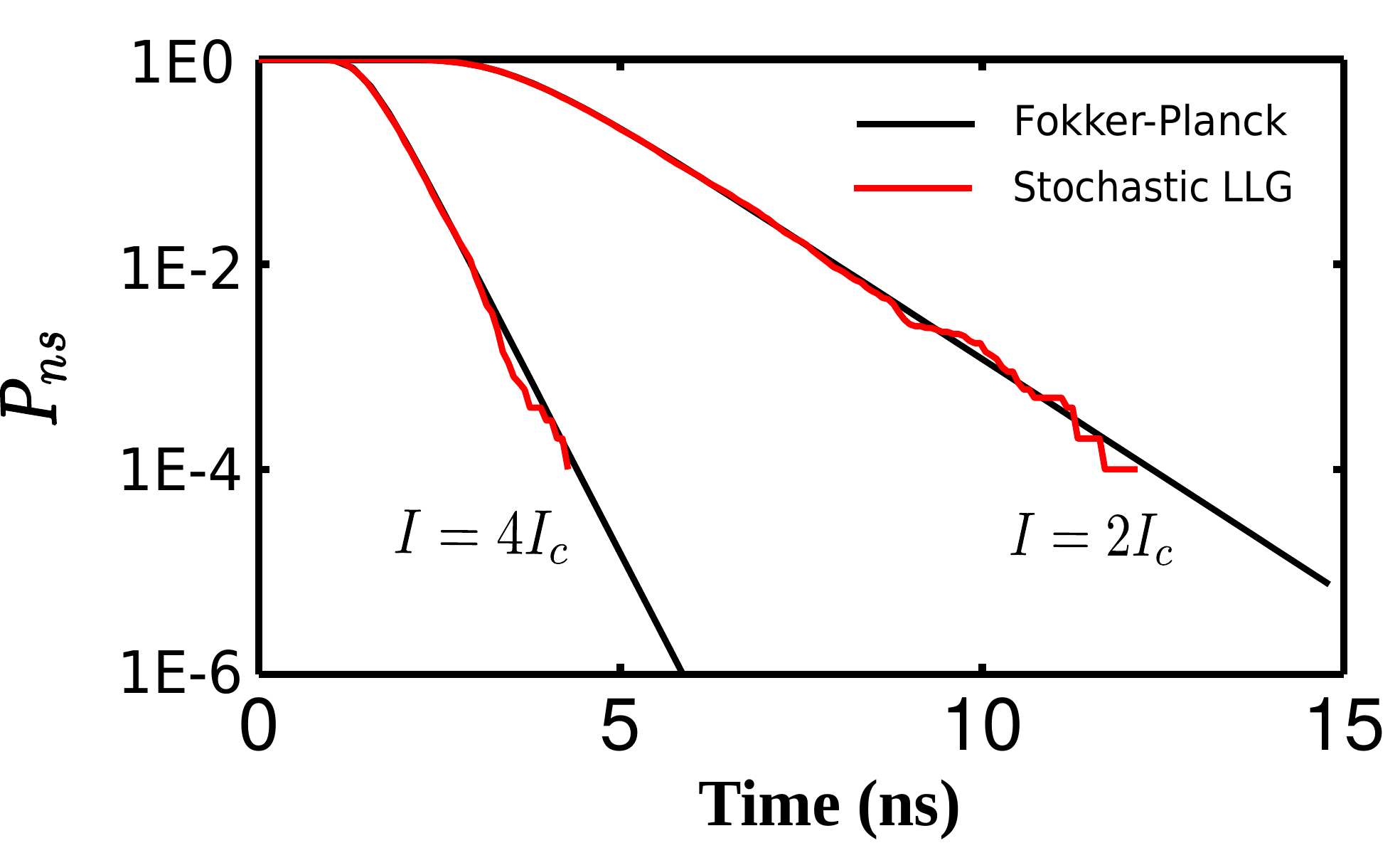}
\caption{Non-switching probability $P_{ns}$ as a function of switching time. Thermal stability $\Delta$ is set as 60. Stochastic LLG result is extracted from an ensemble of 10000 runs.}
\label{methods_compare}
\end{figure}

Here $q$ is the single electron charge and $\mu_B$ is the Bohr magneton. $\mathbf{G}$ is a three dimensional normalized gaussian white noise with $\langle\mathbf{G}\rangle=0$ and $\langle\mathbf{G}^2\rangle=1$. $\mathbf{I}_s$ is the unit vector along the injected spin orientation. To accurately describe switching distribution Eq. \ref{eq:LLGS} needs to be solved for a large number of trials which is computationally expensive. Fig \ref{methods_compare} shows the comparison between the stochastic LLG approach and the upcoming Fokker-Planck approach. The non-switching probability (write error rate WER, defined as the probability of non-switching events when switching is expected) is plotted as a function of switching time. Both simulations give the same result but Stochastic LLG is less efficient (compare 1 shot FPE vs 10000 runs of stochastic LLG).

\subsection{Fokker-Planck Equation}
\subsubsection*{General 2D Fokker-Planck equation}
An alternative way to quantify the statistical nature of STT switching is to solve the corresponding Fokker-Planck equation. The Fokker-Planck method has been applied to describe thermally agitated magnet by Brown\cite{Brown1963}. The method can be generalized to include spin transfer torque. We start from the LLG equation with thermal noise:
\begin{equation}
\frac{\partial \mathbf{m}}{\partial t}=\mathbf{L}\left(\mathbf{m}\right)+\left[\mathbf{m}\times\mathbf{F}_\mathrm{th}\right]
\label{eq:langevin}
\end{equation}
where $\mathbf{L\left(\mathbf{m}\right)}$ includes all the deterministic torques from Eq. \ref{eq:LLGS} (without $\mathbf{h}_\mathrm{th}$ term), with a gaussian distribution of random thermal noise $\mathbf{F}_\mathrm{th}$
\begin{IEEEeqnarray}{rl}
\mathbf{L\left(\mathbf{m}\right)}&=\mathbf{L}_\mathrm{prec}+\mathbf{L}_\mathrm{damp}+\mathbf{L}_\mathrm{STT}\\
\Pi\left(\mathbf{F}_\mathrm{th}\right)&=\frac{1}{\sqrt{8\pi^3D}}\exp{\left(\frac{-\left|\mathbf{F}_\mathrm{th}\right|^2}{2D}\right)}
\label{eq:2DFP1}
\end{IEEEeqnarray}
Instead of keeping track of the random trajectory of $\mathbf{m}$, the Fokker-Plank equation solves for the probability distribution of magnetization:
\begin{equation}
\rho\left(\mathbf{m};t\right) = \int\Pi\left(\mathbf{F}\right)\delta\left(\mathbf{m}-\mathbf{m}_\mathbf{F}\right)d\mathbf{F}
\label{eq:2DFP2}
\end{equation}
From Eq.(\ref{eq:langevin})-(\ref{eq:2DFP2}) one can write down a general Fokker-Planck equation for nano-magnet in the form of a convection-diffusion equation on a 2D spherical surface:
\begin{equation}
\frac{\partial \rho}{\partial t}=-\mathbf{\nabla}\cdot\left(\mathbf{L}\rho\right)+D\mathbf{\nabla}^2\rho
\label{eq:2D_FP}
\end{equation}
where $\rho\left(\theta,\phi;t\right)$ is the probability density of the magnetization in spherical coordinate. The effective `diffusion' constant describing the thermal effect is defined as:
\begin{equation}
D = \frac{\alpha\gamma k_BT}{(1+\alpha^2)\mu_0M_s\Omega}
\end{equation}

\subsubsection*{Reduced 1D Fokker-Planck equation}
For a tunnel junction based on PMA with rotational symmetry, the FPE can be easily reduced to a 1-D differential equation as shown in \cite{Butler2012}:

\begin{equation}
\begin{split}
\frac{\partial\rho(\theta,\tau)}{\partial\tau}=-\frac{1}{\sin\theta}\frac{\partial}{\partial\theta}\left[\sin^2\theta\left(i-h-\cos\theta\right)\rho\left(\theta,\tau\right) \right.\\
\left.-\frac{\sin\theta}{2\Delta}\frac{\partial\rho(\theta,\tau)}{\partial\theta}\right]
\label{eq:1D_FP}
\end{split}
\end{equation}
where $h=H_{z,ext}/H_k$ is the scaled external magnetic field along easy axis. The solution $\rho(\theta,\tau)$ is the probability density of magnetization along $\theta$ at scaled time $\tau$ where $\theta$ is the angle from the easy axis. 

\begin{figure}[H]

 \centering
 \includegraphics[width=7.8cm]{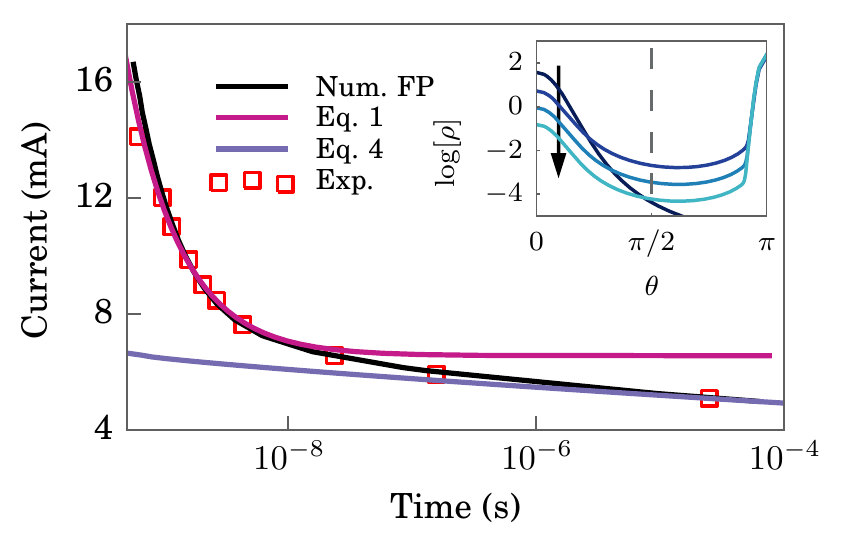}
 \caption{Average switching current ($P_{sw}=0.5$) as a function of pulse width in a 100 nm spin-valve nano-pillar. The experimental data are extracted from Fig. 7(a) of ref \cite{Liu2014}. Inset: Time-evolution of probability distribution at $t=0.1,5.0,10,15\,\mathrm{ns}$ with current $I=6.55\,\mathrm{mA}$. The arrow indicates increasing time.}
\label{fit_ADKent}
\end{figure}

One advantage of numerical method is its universal description throughout all switching regimes. The analytical expressions (Eq. \ref{eq:Sun},\ref{eq:AFP},\ref{eq:thermal},\ref{eq:AFP_thermal}) diverge at the transition of current across the intermediate region. One example is the plot of average switching current as a function of pulse width. The average switching current is defined when the switching probability equals half: $P_{sw}(I_{sw})=0.5$ and this average is usually plotted as a function of pulse width. Fig. \ref{fit_ADKent} shows a $I_{sw}-t$ relation for a $100\,\mathrm{nm}$ diameter metallic spin-valve measured by Bedau {\it et al}\cite{Bedau2010,Liu2014}. The experimental data is fitted by Eq.\ref{eq:Sun} in the large current regime, Eq. \ref{eq:thermal} in the small current regime and Fokker-Planck in all regimes. The physical parameters fitted here are the thermal stability factor $\Delta$, critical current $I_c$ and $\tau_D$. In the case of Eq. \ref{eq:thermal} in the thermal regime, $f_0$ is set to a typical value of $10^9$ Hz. $\beta$ is also treated as a fitting parameter to account for non-idealities. Fitting with Eq.\ref{eq:Sun} and Eq. \ref{eq:thermal} gives $\Delta=63, I_{c}=6.55\ \text{mA},\tau_D=0.26\ \text{ns},\ \beta=1.2$. Fokker-Planck fitting gives $\Delta=80, I_{c}=8.3\ \text{mA},\tau_D=0.25\ \text{ns}$. Given the difference between those two approaches discussed before, this discrepancy is reasonable. Although the switching behavior is not ideal macrospin (see detailed discussions in \cite{Liu2014}), the match is enough to show the applicability of FPE method throughout different regimes.

\section{Write error rate}
In developing energy efficient STTRAMs, a great deal of effort has been devoted into reducing the critical current $I_c$ which is a primary performance factor. In the meantime, practical STT devices demand fairly reliable switching with error rates set by the specific applications. We focus here on write error rate (WER) described earlier. The WER is usually plotted as a function of voltage/current for a given pulse width. Such a plot contains multiple messages. First, the plot indicates the onset of switching which can indicate the average switching current. The plot also sets the approximate boundary for read current since a switching event in read operation causes read error. Second, the lowest error rate achieved at certain current limit will essentially determine the size of integration this device can achieve. Third, when the error rate is plotted in log scale, the slope of the `error tail' contains information about how fast the error rate goes down with increasing current, which characterizes the error margin for the write operation. In general, WER has a complex dependence on current, pulse width and material properties. Current and pulse width are often determined by the application. Therefore we will focus our discussion on a few other related parameters: temperature, current polarization, saturation magnetization, anisotropy and damping. Some parameters are prone to change and hard to control (e.g. temperature rise due to joule heating) while others are often tuned by experimentalists to achieve better performance. We hope to shed some light on how those parameters affect the WER. For the convenience of discussion, we have chosen a set of experimental data from \cite{ikeda2010} and constructed a `reference' device - a perpendicular CoFeB magnetic tunnel junction. The physical parameters are shown in table \ref{tab:tab1}. All the following discussions assume a fixed pulse width of $\tau_{pw}=10\,\mathrm{ns}$. Before going into the discussion, it is worth mentioning that in general specific WER targets need to be met by the memory array or various applications to be a viable product. The obvious way to achieve low WER is to increase the voltage but that comes with a power consumption penalty and more importantly the probability of dielectric breakdown over time. Therefore switching efficiency is another aspect that can help but so far the highest efficiency is achieved in MTJs with smaller diameter \cite{Gajek2012} which is accompanied by very high resistance. In this paper our focus is on the effect of material parameters on the write error and write margin. The actual WER engineering requires diligent handing of all such factors. Nevertheless the approach is quite general and can be easily applied to other discussions.

\begin{table}[H]
\caption {MTJ Parameters} \label{tab:tab1} 
\begin{center}
\begin{tabular}{ |c|c|c| } 
 \hline
 parameter & value & remark\\ \hline
 $d$ & 40 nm & free layer diameter \\ 
 $t$ & 1 nm & free layer thickness \\
 $M_s$ & $1.23\times10^6$A/m & saturation magnetization \\  
 $\alpha$ & 0.027 & magnetic damping\\
 $\Delta$ & 43 & thermal stability \\
 $\eta$ & 0.6 & spin polarization \\
 $RA$ & 18 $\Omega\cdot\mu$m$^2$ & resistance area product \\
 \hline
\end{tabular}
\end{center}
\end{table}

Let us now discuss the parameter dependences on the MTJ WERs, summarized schematically in Fig. \ref{slope}. In the following WER-V plots, the voltages have been scaled by the average switching voltage $V^{ref}_{sw}(\mathrm{WER}=0.5)$ of the reference device to de-emphasize their exact values but to focus on the general trend.

\subsubsection{Temperature}
\begin{figure}
 \centering
 \includegraphics[width=6.5cm]{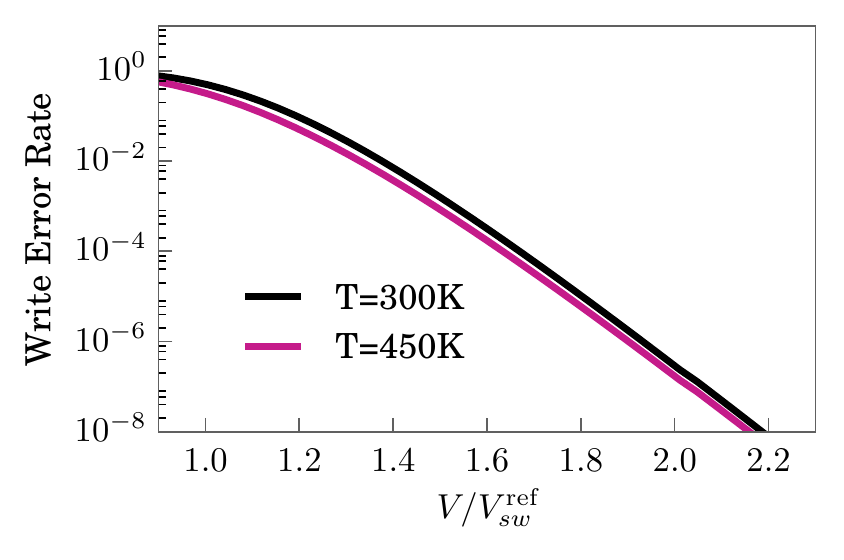}
 \caption{WER as a function of junction bias for different temperatures.}
 \label{Temp_dep} 
\end{figure}

Practical applications of STT are inevitably tied to stochastic switching at finite temperature. Indeed, STTs are prone to environmental change and joule heating during the writing process. The impact of thermal fluctuations on WERs is critical to the operation of STT based devices, the main effects being a change in the angular distribution of initial magnetization and a change in the energy barrier $\Delta=E_b/k_BT$. Fig. \ref{Temp_dep} shows that the impact on WER of a drastic change in temperature is quite minimal. More complicated current driven temperature changes could arise due to self-heating effects that are beyond the scope of this paper \cite{Sousa2004}. Nonetheless, we can estimate their impact on the WER-V curve. While the weak temperature dependence seems counter-intuitive, it is worth emphasizing that a fast write operation usually lies in the current dominated region where thermal effects are limited. For the read process we would expect a larger temperature dependence on the error rate, but since the read current is small the temperature change is limited there as well.

\subsubsection{Spin Polarization}
\begin{figure}
 \centering
 \includegraphics[width=6cm]{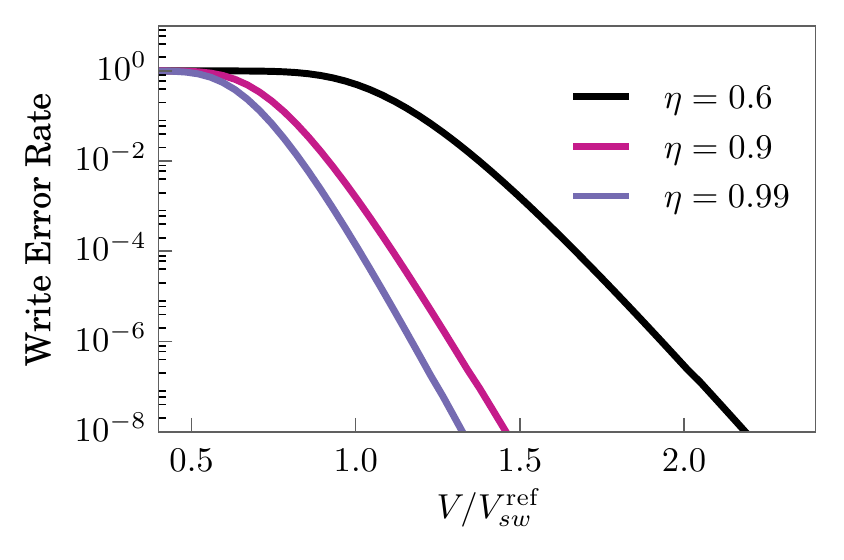}
 \caption{WER as a function of junction bias for various polarization. Higher polarization allows lower critical current and a sharper slope.}
 \label{eta_dep} 
\end{figure}

The degree of spin polarization is a critical determinant of most spintronic applications. It is spin filtering by the fixed magnet that imposes a torque on a noncollinear free magnet. Typically the spin polarization is determined by the materials used but can be largely affected by the interfacial configurations such as symmetry filtering oxides\cite{Butler2001}\cite{Yunkun2016}, defects and strain\cite{Mather2006}\cite{Miao2008}. Fig. \ref{eta_dep} plots the WER-V for various spin polarizations. Understandably, a higher polarization requires a smaller critical current (from Eq. \ref{eq:parameters}) as well as a smaller average switching current ($\mathrm{WER}=0.5$). At the same time, the slope of WER-V increases with spin polarization, meaning a narrower write margin is expected for higher spin polarized contacts. Therefore increasing spin polarization lowers the critical current and reduces noise margin but with $\eta\rightarrow1$ the performance improvement approaches zero.

\subsubsection{Anisotropy, Damping and saturation magnetization}

\begin{figure}
 \centering
 \includegraphics[width=8cm]{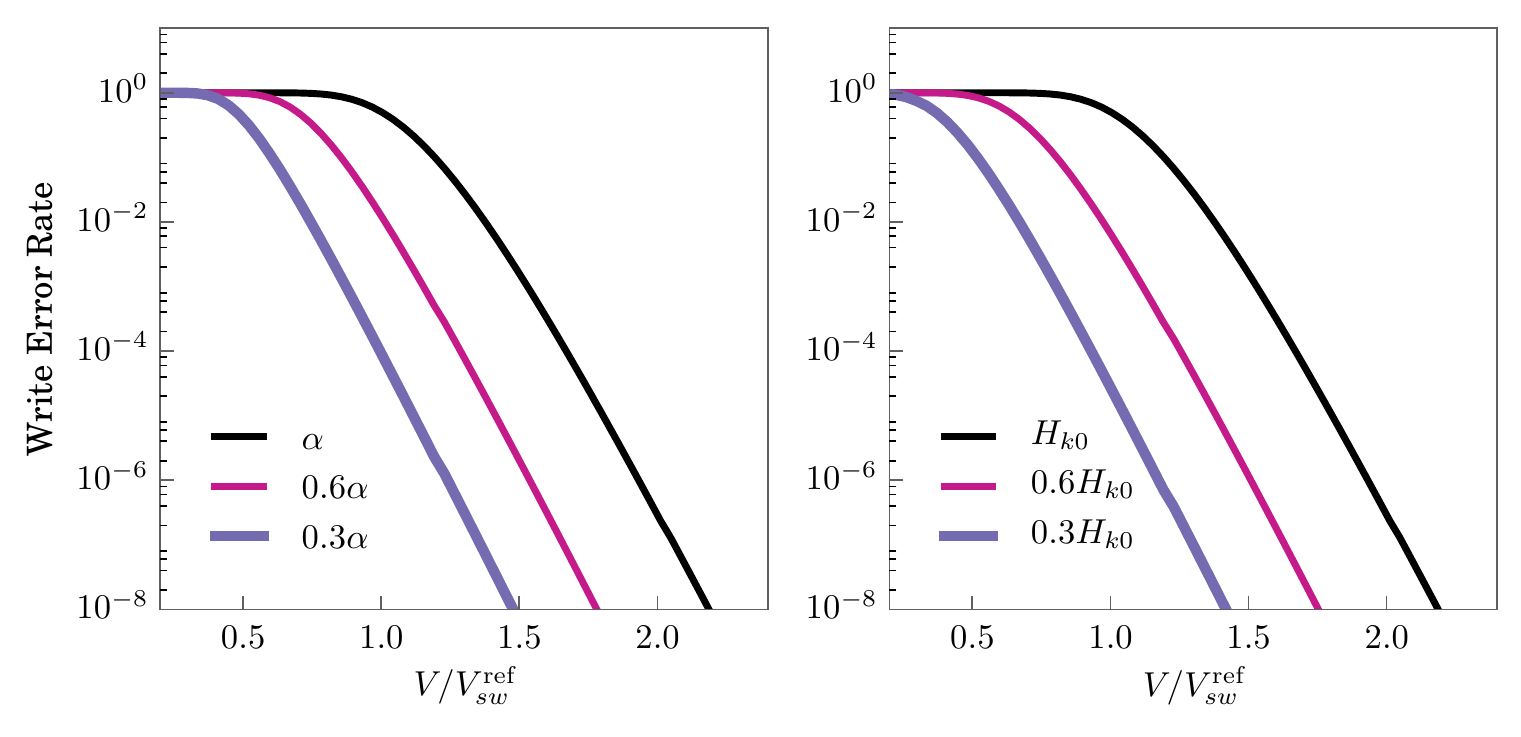}
 \caption{WER-V plot for various magnetic damping and anisotropy. Left: WER-V for different magnetic damping $\alpha$. Right: WER-V for different magnetic anisotropy field $H_k$.}
 \label{Ku_alpha_dep} 
\end{figure}

Besides spin polarization, the other three physical parameters experimentalists often work with are anisotropy, magnetic damping and saturation magnetization. Improvements in fabrication techniques as well as material modeling promise better control over these parameters over time. Fig. \ref{Ku_alpha_dep} shows that magnetic damping $\alpha$ and anisotropy field $H_k$ have almost identical effects on the WER, assuming $M_s$ is held constant. Reducing either parameter would reduce the switching current. This is expected since these are the physical forces that oppose magnetic switching \cite{Sun2000}. Compared to our previous discussion on polarization however, reducing $\alpha$ or $H_k$ does not change the slope of the WER-V curve, meaning the noise margin remains the same even when the average switching current is changed. 

\begin{figure}

 \centering
 \includegraphics[width=8cm]{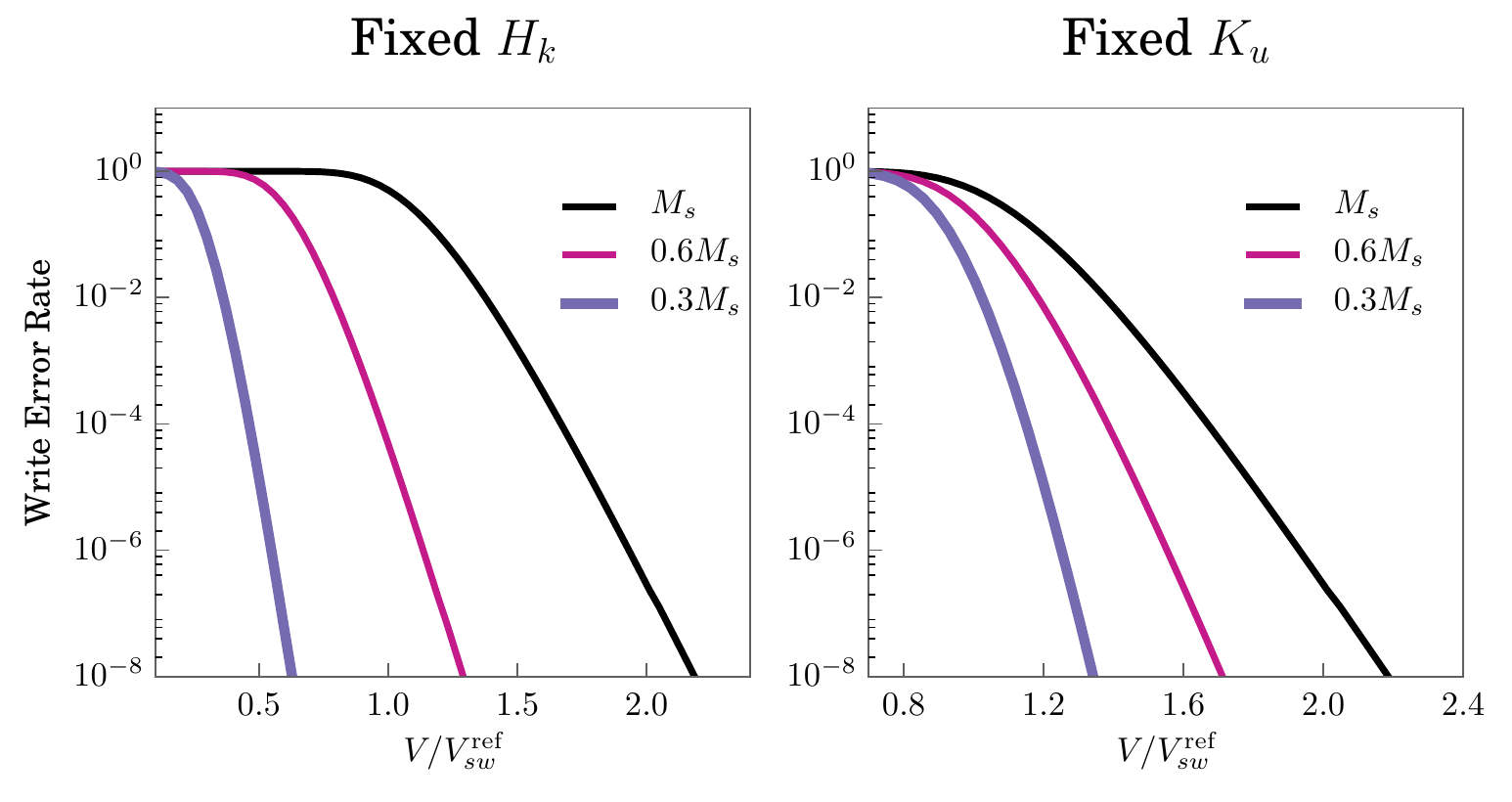}
 \caption{WER-V plot for various saturation magnetization. Left: WER-V with fixed $H_k$. Right: WER-V with fixed $K_u$.}
\label{Ms_dep}
\end{figure}

Fig. \ref{Ms_dep} shows how saturation magnetization $M_s$ affects the WER. Since $M_s$ and $H_k$ are related through Eq. \ref{eq:parameters}, two different scenarios emerge. In the left plot of Fig. \ref{Ms_dep} $H_k$ is kept fixed. This means the overall perpendicular magnetic anisotropy constant $K_u$ changes with $M_s$ (see Eq. \ref{eq:parameters}), so that a reduction in $M_s$ reduces the switching current and increases the slope. In the right plot, $K_u$ is kept unchanged while $H_k$ changes with $M_s$ accordingly. The average switching current is only slightly altered while the slope of WER-V curve changes similar way as the left plot. 
 
\begin{figure}[H]
 \centering
 \includegraphics[width=5cm]{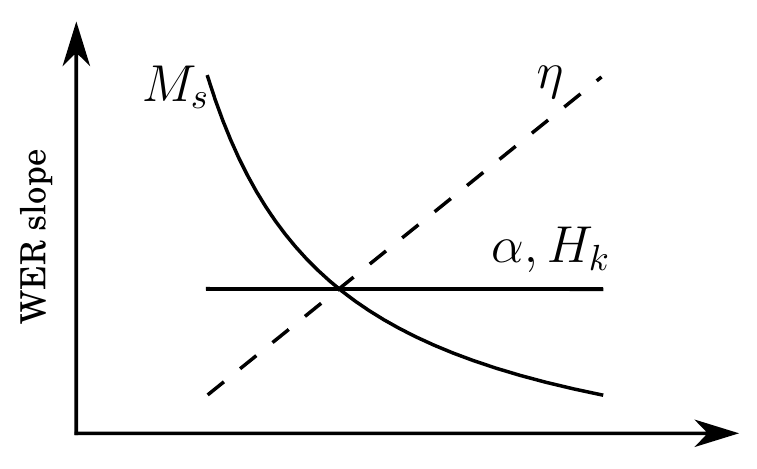}
 \caption{Schematic for the WER slope dependence on various physical parameters.}
\label{slope}
\end{figure}

In the real world it is hard to change one parameter at a time. Indeed, many of the parameters discussed above are in fact, correlated. However, it is not difficult to comprehend their combined effect on the write process. Notice that the average switching current $I_{sw}$ is mostly determined by the intrinsic critical current $I_{c0}$. From Eq. \ref{eq:parameters}, it is easy to see $\alpha,\eta,M_sH_k\propto K_u$ are the factors determining the critical current. The slope of WER-V curve is a function of voltage/current but its asymptotic value at small WER can be approximated from Sun's equation (Eq. \ref{eq:Sun}) in the limit WER$\rightarrow 0$ where we get

\begin{equation}
 \mathcal{S}:=-\frac{d\log[\text{WER}]}{dV}\approx\frac{\mu_B}{M_s\Omega}\frac{\eta}{1+\alpha^2}\frac{t}{qR}
 \label{eq:slope}
\end{equation}
with $R$ the junction resistance. Table \ref{tab:tab2} shows that although the $\mathcal{S}$ value calculated from Eq. \ref{eq:slope} is not always accurate, it correctly captures the overall dependence of WER slope on the physical parameters (Fig. \ref{slope}).

\begin{table}[ht]
\caption {WER slope for different parameters} \label{tab:tab2} 
\centering 
\begin{tabular}{cccccc} %
\toprule
\multirow{2}{*}{
\parbox[c]{.2\linewidth}{\centering param. value}}
  & \multicolumn{2}{c}{decades per 100 mV} &&
\multicolumn{2}{c}{ratio to $\mathcal{S}_0$} \\ 
\cmidrule{2-3} \cmidrule{5-6}

 & {\centering in plot} & {eq. \ref{eq:slope}} && {in plot} & {eq. \ref{eq:slope}}  \\
\midrule
$\mathcal{S}_0$ (ref.) & 1.26  & 1.56  && 1  & 1 \\
$T=450$ & 1.26 & 1.56 && 0.99  & 1 \\
$\eta=0.9$ & 1.92 & 2.35 && 1.52 & 1.5\\  
$\eta=0.99$ & 2.10 & 2.58 && 1.67 & 1.65\\
$0.6\alpha$ & 1.33 & 1.56 && 1.05 & 1 \\
$0.3\alpha$ & 1.33 & 1.56 && 1.05 & 1 \\ 
$0.6H_{k}$ & 1.29 & 1.56  && 1.02 & 1 \\
$0.3H_{k}$ & 1.32 & 1.56  && 1.05 & 1 \\
$0.6M_{s}$ & 2.09 & 2.61  && 1.66 & 1.67 \\
$0.3M_{s}$ & 4.10 & 5.22  && 3.25 & 3.34 \\
\bottomrule
\end{tabular}
\end{table}

\section{2D Fokker-Planck for non-symmetric system}
For a typical perpendicular MTJ with cylindrical symmetry, reduced FPE (Eq. \ref{eq:1D_FP}) suffices to describe the stochastic nature of STT switching. In other cases where cylindrical symmetry is broken, such as canted magnet for example, the more general version in Eq. \ref{eq:2D_FP} should be used. In the absence of added symmetries, this must be solved numerically. We solve Eq. \ref{eq:2D_FP} through finite-element method with triangular meshes generated on a unit spherical surface \cite{zhang2001}\cite{persson2004}. The differential operator is discretized with Galerkin's method and the time evolution is calculated through Crank Nicolson's method. The solution is a 2-dimensional probability density $\rho(\theta,\phi;t)$ evolving on the surface of a unit sphere. The WER is then evaluated as the total integrated probability lingering on the upper hemisphere:

\begin{equation}
\mathrm{WER}=\int_0^{\pi/2}\int_0^{2\pi}\rho\left(\theta,\phi;t\right)d\phi d\theta
\end{equation}

\begin{figure}
 \centering
 \includegraphics[width=8cm]{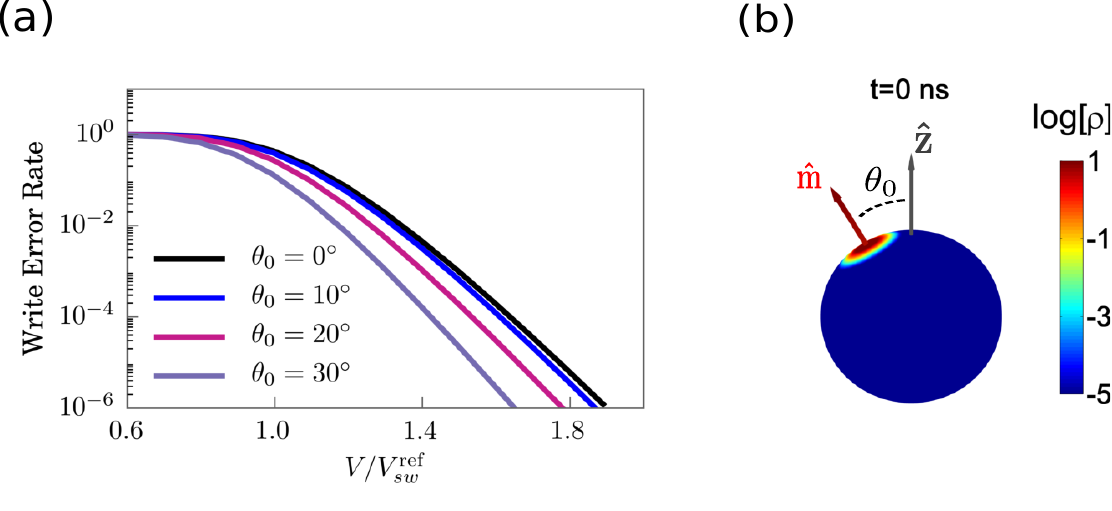}
  \includegraphics[width=7cm]{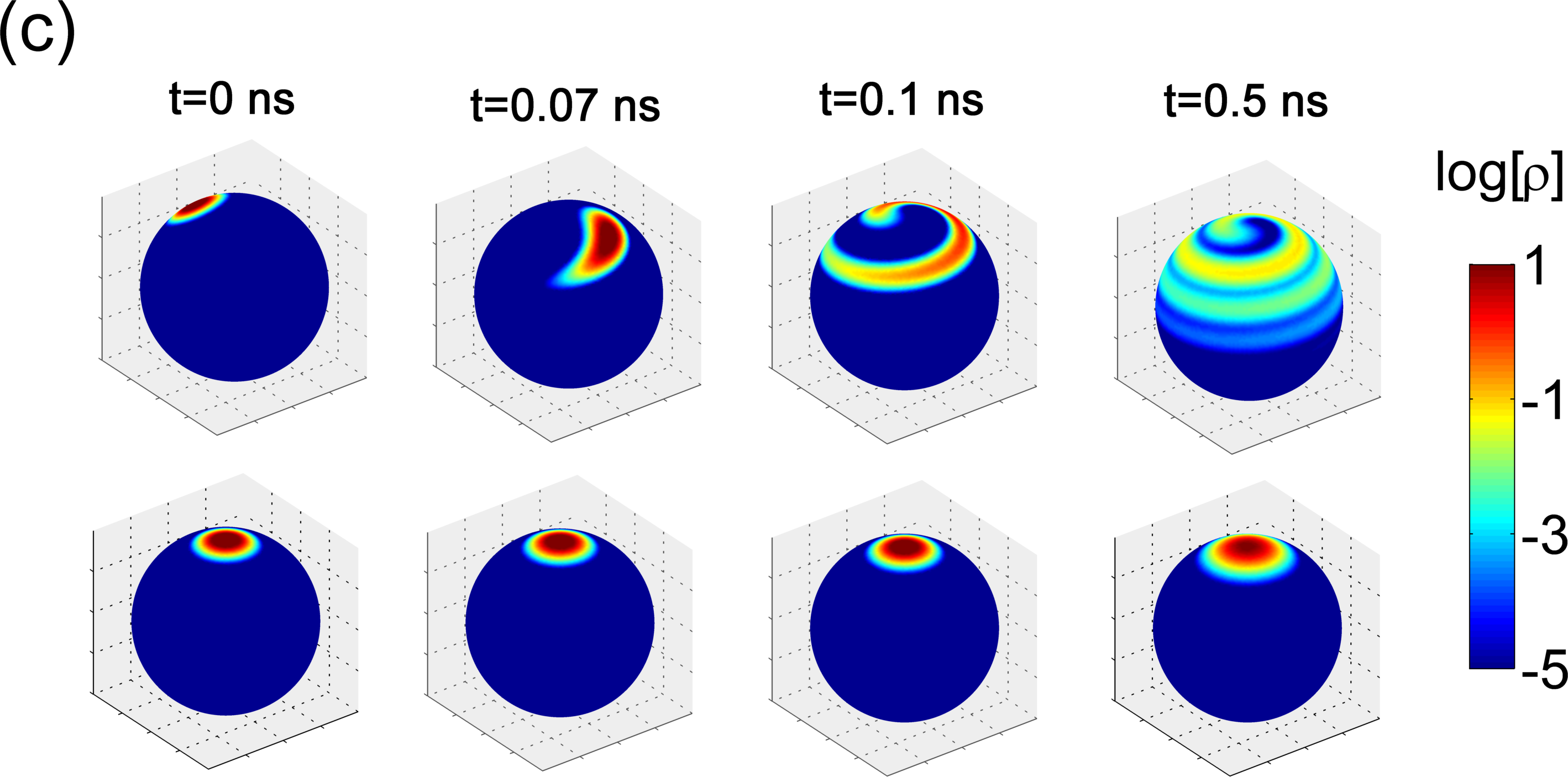}
 \caption{(a) The write error rate of magnetization switching from an initial angle. (b) Initial distribution of magnetization probability $\rho(\theta,\phi)$ on a unit sphere. (c) Time evolution of the probability density for canted case (top) with $\theta_0=30^{\circ}$ and collinear case (bottom) with $\theta_0=0^{\circ}$ at $1.25V^{ref}_{sw}$ junction bias.}
\label{Tilt}
\end{figure}

The reason we need a general 2D Fokker-Planck solver is because the initial incubation phase of STT switching cannot be simply overcome by adjusting material parameters discussed so far. To reduce switching delay and error rates, efficient hybrid switching schemes have been proposed to bypass the initial stagnation period where the fixed and free magnets are collinear. These schemes include a thermal torque applied to excite the magnetization to a larger angle to facilitate switching \cite{mojumder2012}, as well as an in-plane spin-orbital torque arising from a Spin Hall effect to disturb the initial magnetization \cite{van2014}. To capture those dynamics we have to go beyond 1D FPE. We leave the details on magnetization switching and subsequent dynamics to future studies and focus here on the numerical outcomes of the 2D FPE. Fig. \ref{Tilt} shows the WER of  our reference system but with a `tilted' initial angle. The assumption is that the initial magnetization still obeys a Boltzmann distribution except the maximum probability density is inclined at an angle $\theta_0$ with respect to the $z$ axis (easy axis) shown in Fig. \ref{Tilt}(b).  Fig. \ref{Tilt}(a) shows that the initial excitation (usually fast, $<1\mathrm{ns}$, so any overhead delays can be neglected) helps reduce both the critical write current and write noise margin. Fig. \ref{Tilt} (c) further illustrates the different types of switching behavior in a canted system versus a collinear system. Where as in the collinear case, the initial magnetization distribution remains almost unchanged for some time and starts to `diffuse' to the -z direction, in the canted case the magnetization starts to precess and move to the -z direction almost immediately after the current is applied. Therefore the initial delay in the collinear case is avoided in a canted system. We should caution that our assumption of the Boltzmann distribution of the tilted initial angle is likely over-simplified. A full simulation involving the initial excitation process is necessary to see how the switching dynamics plays out in these schemes. This is possible with the 2D FPE numerical tool in its general form (Eq. \ref{eq:2D_FP}), which can easily incorporate different torques from external magnetic fields or spin-orbital couplings. The purpose here is to show a fast numerical result based on the 2D FPE with a simplified assumption, relegating further details to future publications.

\section{Conclusion}
In summary, we have discussed thermal noise induced write error rates in PMA systems with a numerical simulation of the Fokker-Planck equation. The effects of several material parameters on write error are investigated. Among them spin polarization and saturation magnetization change both average switching current and write error margin, while magnetic damping and anisotropy field only affect the switching current. A more general 2D Fokker-Planck solver is needed for geometries beyond PMA or hybrid non-collinear switching schemes with broken cylindrical symmetry.


%



\section*{Acknowledgment}
We acknowledge generous supported from the NSF Grant No. CCF1514219. We also would like to thank Prof. W. H. Butler and P. Visscher from University of Alabama for discussions on the Fokker-Planck methods. The authors also thank Dr. Kangho Lee for valuable discussions on STT-MRAM write error.

\ifCLASSOPTIONcaptionsoff
  \newpage
\fi



%



\bibliography{FP_bib}
\bibliographystyle{IEEEtran}
%

\begin{IEEEbiography}
[{\includegraphics[width=1in,height=1.25in,clip,keepaspectratio]{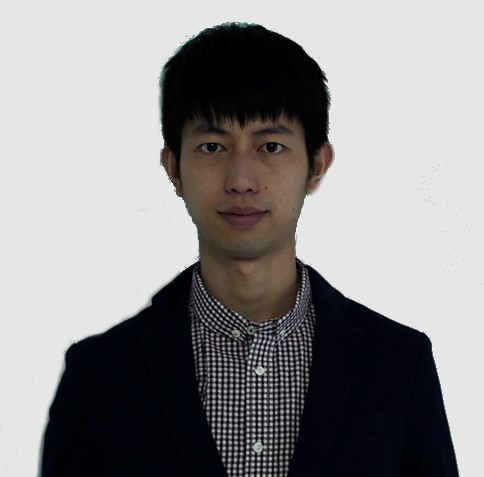}}]{Yunkun Xie}
Yunkun Xie received the B.S degree in Physics from Peking University, Beijing, China, in in 2012. He is currently working toward the Ph.D. degree in electrical engineering at University of Virginia, Charlottesville, VA. His current research interests include first principle material modeling of transport on magnetic tunnel junctions, device simulation of nano-magnetic memory/logic including Spin-Transfer Torque RAM (STT-RAM) and multiferroics. 
\end{IEEEbiography}

\begin{IEEEbiography}[{\includegraphics[width=1in,height=1.25in,clip,keepaspectratio]{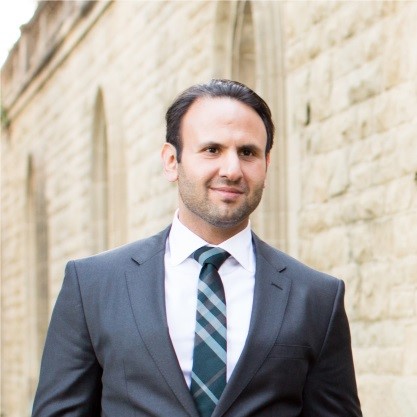}}]{Behtash Behin-Aein}
Behtash Behin-Aein is currently with GLOBALFOUNDRIES. His interests include STT-MRAM, spin based logic, belief networks and combinatorial optimization. He earned his PhD in Electrical and Computer Engineering from Purdue University in 2010. He is an author of more than 35 research paper, book chapters and patents with more than 750 citations and has given 11 contributed and 9 invited talks.  He is the recipient of the 2012 Proctor prize grant-in-aid of research from scientific research society Sigma XI also featured in American Scientist. Behin-Aein has been a lead industry liaison for Semiconductor Research Corporation (SRC) on spin based devices, lead principal investigator for Department of Energys ALCC award, distinguished industry associate for C-SPIN and an invited author for MRS bulletin. He has served as chairperson and judge in international magnetics and SRC TECHCON conferences. 
\end{IEEEbiography}

\begin{IEEEbiography}
[{\includegraphics[width=1in,height=1.25in,clip,keepaspectratio]{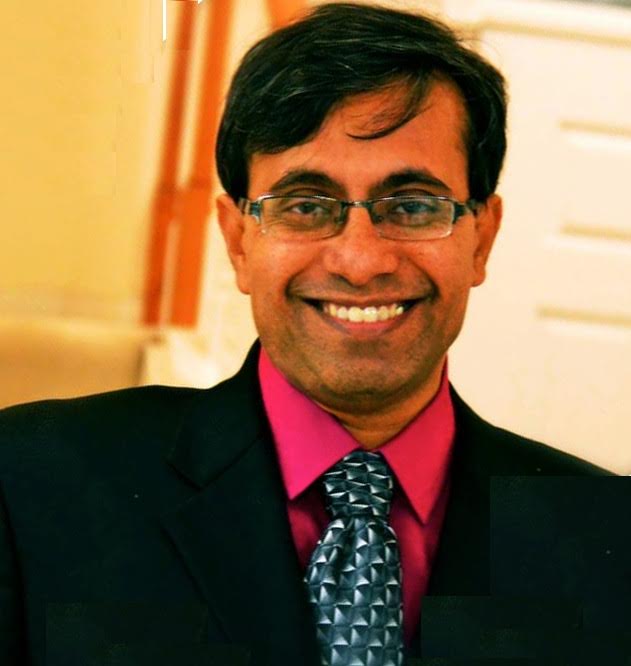}}]{Avik Ghosh}
Avik Ghosh is Professor of Electrical and Computer Engineering and Department of Physics at
the University of Virginia. He has over 90 refereed papers and
book chapters and 2 upcoming books in the areas of computational 
nano-electronics and low power devices including 2D materials,
molecular electronics, subthermal switching, nanomagnetic materials and devices, and nanoscale heat flow. He completed his PhD in physics from the Ohio State University and Postdoctoral Fellowship in Electrical Engineering at Purdue University. He is a Fellow of the Institute of Physics (IOP), senior member of the IEEE, and has received the IBM Faculty Award, the NSF CAREER Award, a best paper award from the Army Research Office, the Charles Brown New Faculty Teaching Award and UVa's All University Teaching Award.
\end{IEEEbiography}





\end{document}